\documentclass[conference]{IEEEtran}
\IEEEoverridecommandlockouts
\usepackage{cite}
\usepackage{amsmath,amssymb,amsfonts}
\usepackage{algorithmic}
\usepackage{graphicx}
\usepackage{textcomp}
\usepackage{xcolor}
\def\BibTeX{{\rm B\kern-.05em{\sc i\kern-.025em b}\kern-.08em
    T\kern-.1667em\lower.7ex\hbox{E}\kern-.125emX}}

\def\c1{{\textcircled{a}}}
\def\ba{{\mathbf{a}}}
\def\bb{{\mathbf{b}}}

\def\bh{{\mathbf{h}}}

\def\bm{{\mathbf{m}}}
\def\bn{{\mathbf{n}}}

\def\bu{{\mathbf{u}}}
\def\bv{{\mathbf{v}}}

\def\bx{{\mathbf{x}}}
\def\by{{\mathbf{y}}}

\def\bA{{\mathbf{A}}}
\def\bB{{\mathbf{B}}}
\def\bC{{\mathbf{C}}}
\def\bD{{\mathbf{D}}}
\def\bE{{\mathbf{E}}}

\def\bH{{\mathbf{H}}}
\def\bI{{\mathbf{I}}}

\def\bN{{\mathbf{N}}}
\def\bM{{\mathbf{M}}}

\def\bP{{\mathbf{P}}}
\def\bQ{{\mathbf{Q}}}
\def\bR{{\mathbf{R}}}

\def\bT{{\mathbf{T}}}

\def\bW{{\mathbf{W}}}
\def\bX{{\mathbf{X}}}
\def\bY{{\mathbf{Y}}}
\def\bZ{{\mathbf{Z}}}

\def\tr{{\textrm{{tr}}}}
\def\T{{\top}}
\renewcommand\Re{{\textrm{Re}}}
\renewcommand\vec{{\textrm{vec}}}
\def\HH{{\dagger}}

\begin{document}

\title{A Probabilistic Model-Based Robust Waveform Design for MIMO Radar Detection\\
%\thanks{This work was supported in part by the National Natural Science Foundation of China under Grant 61671453 and 61801500, in part by the Young Elite Scientist Sponsorship Program under Grant 17-JCJQ-QT-041, and in part by the Anhui Provincial Natural Science Foundation under Grant 1908085QF252.}
}

\author{\IEEEauthorblockN{Xuyang Wang}
\IEEEauthorblockA{\textit{College of Electronic Engineering} \\
\textit{National University of Defense Technology}\\
Hefei, China \\
wangxuyang@nudt.edu.cn}
\and
\IEEEauthorblockN{Bo Tang}
\IEEEauthorblockA{\textit{College of Electronic Engineering} \\
\textit{National University of Defense Technology}\\
Hefei, China \\
tangbo06@gmail.com}
\and
\IEEEauthorblockN{Ming Zhang}
\IEEEauthorblockA{\textit{College of Electronic Engineering} \\
\textit{\small National University of Defense Technology}\\
Hefei, China \\
lgdxzm@sina.com}

}

\maketitle

\begin{abstract}
This paper addresses robust waveform design for multiple-input-multiple-output (MIMO) radar detection.  A probabilistic model is proposed to describe the target uncertainty. Considering that waveform design based on maximizing the probability of detection is intractable, the relative entropy between the distributions of the observations under two hypotheses (viz., the target is present/absent) is employed as the design metric. To tackle the resulting non-convex optimization problem, an efficient algorithm based on minorization-maximization (MM) is derived. Numerical results demonstrate that the waveform synthesized by the proposed algorithm is more robust to model mismatches.
%In this paper, we proposed a waveform design algorithm for multiple-input-multiple-output (MIMO) radar target detection. Compared with the early method, the proposed method has robust performance. In the past, we have proposed optimal waveforms for Swerling 0/deterministic target detection. But once real target deviates from prior knowledge, the performance will decline. Therefore, in this paper, we proposed a new robust method, which has a better performance in target detection. And numerical examples have proved that the proposed algorithm can obtain higher relative entropy and get a superior performance of probability of detection.
\end{abstract}

\begin{IEEEkeywords}
MIMO radar, robust waveform design,  probabilistic model, minorization-maximization (MM), relative entropy.
\end{IEEEkeywords}

\section{Introduction}
In recent years, multiple-input-multiple-output (MIMO) radar has gained considerable attentions because of its superior performance \cite{LiMIMObook2008}. Generally, MIMO radar can be divided into two categories. One is called statistical MIMO radar (or distributed MIMO radar), i.e., MIMO radar systems with widely separated antennas \cite{HaimovichSeparated2008}. The spatial diversity provided by statistical MIMO radar allows to improve the detection performance of a fluctuating target. The other type is called colocated MIMO radar, whose antennas are closed to each other \cite{LiColocated2007}. Different from the phased-array radar,  the antennas of colocated MIMO radar can emit different waveforms. The additional waveform diversity of colocated MIMO radar enables better target detection performance in the presence of interference and improved parameter identifiability.

For both types of MIMO radar, it is important to design the waveforms appropriately. In the past years, many criteria have been adopted to design MIMO radar waveforms, including minimizing the auto-correlations and cross-correlations of the waveforms (see, e.g., \cite{HeCorrelations2009,HeWaveformBook2012} and the references therein), maximizing the signal-to-interference-plus-noise ratio (SINR) \cite{ChenExtended2009,TangJoint2016,Cui2017SpaceTime,Tang2020Polyphase}, maximizing the mutual information between the receive signals and the target response \cite{MaioPrinciples2007,Tang2010IT,Tang2019Spectrally}, to name just a few.

%In \cite{HeCorrelations2009}-\cite{HeWaveformBook2012}, the authors tried to minimize the auto-correlation and cross-correlation of the waveforms. In \cite{ChenExtended2009}-\cite{TangJoint2016}, the authors focused on maximizing signal-to-interference-plus-noise ratio (SINR) to design waveforms. In \cite{MaioPrinciples2007}, the authors designed waveforms using the metric of lower Chernoff bound.

In this paper, we consider waveform design for MIMO radar detection. Different from the previous studies  \cite{Tang2010IT,TangEntropy2015,Tang2021RangeSpread}, we take account of the model uncertainty about the target (due to estimation errors, inaccurate prior knowledge, etc.). A probabilistic model is proposed to describe the target uncertainty. Then a hypothesis testing is established for the proposed target detection problem. Given that the optimization of waveforms based on maximizing the probability of detection is intractable, we resort to an information-theoretic approach to design the waveforms. We devise a minorization-maximization (MM) based algorithm to tackle the non-convex waveform design problem. Results are provided to show the robustness of the proposed algorithm.

\section{Problem Formulation}
Consider a MIMO radar with $N_\textrm{T}$ transmit antennas and $N_\textrm{R}$ receive antennas. Let $\bx_m$ denote the waveform transmitted by the $m$th transmitter and let $L$ denote its code length. Similar to \cite{Tang2010IT,TangEntropy2015,Tang2016Rician}, we consider a unified signal model, which can be written as
\begin{equation} \label{eq:SigModel}
  \bY = \bX\bH + \bN,
\end{equation}
where $\bY \in \mathbb{C}^{L\times N_\textrm{R}}$ denotes the received signals, $\bX=[\bx_1,\bx_2,\ldots,\bx_{N_\textrm{T}}]\in \mathbb{C}^{L\times N_\textrm{T}}$ is the waveform matrix, $\bH\in \mathbb{C}^{N_\textrm{T}\times N_\textrm{R}}$ denotes the target response (the $(m,n)$th element of $\bH$ stands for the response from the $m$th transmitter to the $n$th receiver), and $\bN$ is the receiver noise. As shown in \cite{TangMinorization2018}, the signal model in \eqref{eq:SigModel} can be used for various types of MIMO radar, including the colocated MIMO radar and distributed MIMO radar. For example, if we consider a colocated MIMO radar, then the target response can be written as
\begin{equation}%\label{}
  \bH = \sum\nolimits_{k} \alpha_{t,k} \ba(\theta_{t,k})\bb^\T(\theta_{t,k}),
\end{equation}
where $\alpha_{t,k}$, $\theta_{t,k}$, $\ba(\theta_{t,k})$, and $\bb(\theta_{t,k})$ are the amplitude, the direction of arrival (DOA), the transmit array steering vector, and the receive array steering vector of the $k$th  target, respectively.

In this paper, we focus on the design of waveforms to enhance the detection performance of MIMO radar systems. Note that the detection performance of radar systems is closely related to the signal-to-noise ratio (SNR), which can be defined as follows:
\begin{equation}\label{eq:SNR}
  \texttt{SNR} = \frac{\tr(\bX\bH \bH^\HH\bX^\HH)}{\textrm{E}(\tr(\bN\bN^\HH))}.
\end{equation}
Assume that the receiver noise is white, then $\textrm{E}(\tr(\bN\bN^\HH)) = LN_\textrm{R} \sigma^2$, where $\sigma^2$ is the power of the noise. Therefore, to improve the target detection performance, we can maximize the SNR via the design of waveforms. The associated waveform design problem can be formulated as
\begin{equation}\label{eq:NominalDesign}
  \max_{\bX} \tr(\bX\bH \bH^\HH\bX^\HH), \textrm{s.t.}\  \textrm{tr}(\bX\bX^\HH)\leq P_t,
\end{equation}
where $P_t$ denotes the total available transmit energy. It is well-known that the optimization problem in \eqref{eq:NominalDesign} admits a closed-form solution (see, e.g., \cite{Tang2016Rician} for the details):
\begin{equation} \label{eq:NominalSolution}
\bX=\sqrt{P_t}\bu_x\bv_{d,1}^\HH,
\end{equation}
where $\bu_x\in \mathbb{C}^{L\times 1}$ is an arbitrary normalized vector, $\bv_{d,1}\in \mathbb{C}^{N_\textrm{T} \times 1}$ is the eigenvector associated with the largest eigenvalue of $\bR_t=\bH\bH^\HH$. Note that to obtain the optimal solution in \eqref{eq:NominalSolution}, the target response matrix $\bH$ should be known \emph{a priori}. However, due to estimation errors, the estimated target response might be inaccurate. As a result, when the transmit waveforms are designed based on \eqref{eq:NominalSolution}, the performance of MIMO radar system might degrade.

To improve the target detection performance in the presence of estimation errors, we consider the robust design of waveforms. To this end, we define $\bh = \vec(\bH)$. We consider a probabilistic model for $\bh$ and assume that $\bh$ is circularly-symmetric Gaussian, with mean $\bh_d$ and covariance matrix $\bR_\textrm{H}$, respectively. It is worth noting that $\bh_d$ can be obtained by the available prior knowledge $\bH$, and $\bR_\textrm{H}$, which is used to rule the uncertainty, can be obtained by the data from previous scans or specified by the user.

Next we establish the following binary hypothesis test for the target detection problem:
\begin{equation}
\left\{
\begin{aligned}
\mathcal{H}_0:& \by=\bn, \\%why not center duiqi?
\mathcal{H}_1:& \by= \widetilde{\bX}\bh+\bn,
\end{aligned}
\right.
\end{equation}
where $\by = \vec(\bY)$, $\bn = \vec(\bN)$, $\widetilde{\bX} = (\bI_{N_\textrm{R}}\otimes \bX)$, and we have used the fact that $\vec(\bA\bB\bC) = (\bC^\T\otimes \bA) \vec(\bB)$.
%$\by\in \mathbb{C}^{LN_R\times 1}$ denotes the received signal of MIMO radar and $\bn\in\mathbb{C}^{LN_R\times 1}$ is the receiver noise. Then we assume $\bn$ is circularly-symmetric white Gaussian, with power equal to $\sigma^2$.

Note that the probability density function (PDF) of $\by$ under $\mathcal{H}_0$ is given by
\begin{equation*}
P_0(\by)=\frac{1}{\pi^{LN_\textrm{R}}\sigma^{2LN_\textrm{R}}}\exp(-\by^\HH\by/\sigma^2),
\end{equation*}
and under $\mathcal{H}_1$, the PDF of $\by$ is given by
\begin{align*}
P_1(\by)=\frac{1}{\pi^{N_\textrm{RL}}\det(\bR_1)}\exp[-(\by-\widetilde{\bX}\bh_d)^\HH\bR_1^{-1}(\by-\widetilde{\bX}\bh_d)],
\end{align*}
where
\begin{equation}\label{eq:R1}
  \bR_1 = \widetilde{\bX}\bR_\textrm{H}\widetilde{\bX}^\HH+\sigma^2\bI_{N_\textrm{RL}},
\end{equation}
and $N_\textrm{RL}=LN_\textrm{R}$.
Therefore, the Neyman-Pearson (NP) detector\cite{Kaystatistical1998} decides $\mathcal{H}_1$ if
\begin{align} \label{eq:NPdetector}
&\by^\HH(\bI - \sigma^2\bR_1^{-1})\by+2\sigma^2\Re(\by^\HH\bR_1^{-1}\widetilde{\bX}\bh_d)>\gamma,
\end{align}
where $\gamma$ is the detection threshold.

One straightforward way to design the robust waveform is to analyze the probability of detection for the NP detector in \eqref{eq:NPdetector} first (given the probability of false alarm), and optimize the waveform based on maximizing the probability of detection. However, the probability of detection associated with \eqref{eq:NPdetector} is too complex to be used as a design metric. Alternatively, we resort to relative entropy to design the waveforms. Indeed, Stein's lemma states that a larger relative entropy can result in a higher probability of detection asymptotically. Therefore, to improve the target detection performance of MIMO radar systems, we aim to design waveforms to maximize the relative entropy.

%Same as \cite{RN33}, we adopt relative entropy as the principle to design our waveforms. Stein's lemma proves that if we have a larger relative entropy, we will have a larger probability of detection. Therefore, we give the expression of relative entropy first, then we will solve the optimization problem to obtain the optimal waveforms we need. The relative entropy is given by
The relative entropy between $P_0{(\by})$ and $P_1{(\by})$ is given by
\begin{align}
\textrm{D}(P_0||P_1)
=&\int P_0(\by)\log\frac{P_0{(\by})}{P_1{(\by})}d\by \nonumber \\
=&\log\det(\bR_1) + \tr(\bR_1^{-1}(\widetilde{\bX}\bh_d\bh_d^\HH\widetilde{\bX}^\HH+\sigma^2\bI_{N_\textrm{RL}})) \nonumber \\
  &-LN_\textrm{R}(1+\log\sigma^2).
\end{align}

%Then we can obtain
%\begin{align}
%&D(P_0||P_1)=\log\det(\widetilde{\bX}\bR_H\widetilde{\bX}^H+\sigma^2\bI_{LN_R})
%\nonumber\\
%&+E((\by-\widetilde{\bX}\bh_d)^H(\widetilde{\bX}\bR_H\widetilde{\bX}^H+\sigma^2\bI_{LN_R})^{-1}(\by-\widetilde{\bX}\bh_d)|\mathcal{H}_0) \nonumber\\
%&-E(\by^H\by/\sigma^2|\mathcal{H}_0)-LN_R\log\sigma^2.
%\end{align}
%where E$(\cdot|\mathcal{H}_0)$ denotes the expectation of a random variable under hypothesis $\mathcal{H}_0$.

%According to \cite{RN33}, we can obtain
%\begin{align}
%&D(P_0||P_1)=\log\det(\widetilde{\bX}\bR_H\widetilde{\bX}^H+\sigma^2\bI_{LN_R})
%\nonumber \\
%&+\textrm{tr}[(\widetilde{\bX}\bR_H\widetilde{\bX}^H+\sigma^2\bI)^{-1}(\widetilde{\bX}\bh_d\bh_d^H\widetilde{\bX}^H+\sigma^2\bI_{LN_R})]
%\nonumber \\
%&-LN_R(1+\log\sigma^2).
%\end{align}

%Then the optimal problem can be given by
Then the waveform design problem based on maximizing relative entropy can be formulated by
\begin{align} \label{eq:RobustDesign}
\max\limits_{\bX} & \ \log\det(\bR_1) + \tr(\bR_1^{-1}(\widetilde{\bX}\bh_d\bh_d^\HH\widetilde{\bX}^\HH+\sigma^2\bI_{N_\textrm{RL}}))\nonumber\\
%\log\det(\widetilde{\bX}\bR_H\widetilde{\bX}^H+\sigma^2\bI_{LN_R})+
%\nonumber\\
%&\textrm{tr}[(\widetilde{\bX}\bR_H\widetilde{\bX}^H+\sigma^2\bI)^{-1}\widetilde{\bX}\bh_d\bh_d^H\widetilde{\bX}^H]+\sigma^2\textrm{tr}(\widetilde{\bX}\bR_H\widetilde{\bX}^H+\sigma^2\bI)
%\nonumber\\
\textrm{s.t.} &\ \textrm{tr}(\bX\bX^\HH)\leq P_t.
\end{align}

It can be checked that the optimization problem is non-convex and difficult to solve. In the following, we develop an efficient algorithm based on MM to tackle the optimization problem in \eqref{eq:RobustDesign}.
%Then we can find this problem is a non-convex optimization problem, which is difficult to solve. Therefore, we need to develop an efficient method to solve this problem.
%\IEEEpubidadjcol

\section{Algorithm Design}
%In this paper, we proposed an algorithm based on MM method. MM method is a useful method to solve non-convex optimization problem, and it is widely used in many problems. Using MM method to solve non-convex problem, we need to find an objective function first.

For simplicity we assume that the power of noise is $\sigma^2=1$. Let $\bR_{X}=\bX^\HH\bX$ denote the waveform covariance matrix, and $\widetilde{\bR}_{X}=\widetilde{\bX}^\HH\widetilde{\bX}=\bI_{N_\textrm{R}}\otimes \bR_{X}$.
Using the standard property of matrix determinant that $\det(\bI+\bA\bB)=\det(\bI+\bB\bA)$, we can rewrite $\log\det(\bR_1)$ as
\begin{center}
\begin{math}
\log\det(\bR_1)=\log\det(\bR_\textrm{H}^{\frac{1}{2}}\widetilde{\bR}_{X}\bR_\textrm{H}^{\frac{1}{2}}+\bI).
\end{math}
\end{center}

Then the objective of \eqref{eq:RobustDesign} can be divided into three parts:
\begin{align} \label{eq:fX}
f(\bX)=
&\underbrace{\log\det(\bR_\textrm{H}^{\frac{1}{2}}\widetilde{\bR}_{X}\bR_\textrm{H}^{\frac{1}{2}}+\bI)}_{\textrm{Part I}}+\underbrace{\textrm{tr}(\bR_1^{-1}\widetilde{\bX}\bh_d\bh_d^\HH\widetilde{\bX}^\HH)}_{\textrm{Part II}}
\nonumber \\
&+\underbrace{\textrm{tr}(\bR_1^{-1})}_{\textrm{Part III}}.
\end{align}

The key step of MM methods is to find a surrogate function (i.e., a minorizer) $Q(\bX;\bX_k)$, which satisfies:
\begin{subequations}
  \begin{align}%\label{}
  Q(\bX;\bX_k) & \leq f(\bX), \\
  Q(\bX_k;\bX_k) & = f(\bX_k).
\end{align}
\end{subequations}

To this purpose, next we construct minorizers for each part of \eqref{eq:fX}, respectively.
\subsection{Minorizing Part I}
By using the standard property of matrix determinant, we have
\begin{center}
\begin{math}
\log\det(\bR_\textrm{H}^{\frac{1}{2}}\widetilde{\bR}_{X}\bR_\textrm{H}^{\frac{1}{2}}+\bI)=-\log\det(\bR_\textrm{H}^{\frac{1}{2}}\widetilde{\bR}_{X}\bR_\textrm{H}^{\frac{1}{2}}+\bI)^{-1}.
\end{math}
\end{center}

According to  the matrix inversion lemma \cite{HornMatrix1990}, we obtain
\begin{align} \label{eq:Inversionmatrix}
(\bR_\textrm{H}^{\frac{1}{2}}\widetilde{\bR}_{X}\bR_\textrm{H}^{\frac{1}{2}}+\bI)^{-1}=\bI-\bR_\textrm{H}^{\frac{1}{2}}\widetilde{\bX}^\HH\bR_1^{-1}\widetilde{\bX}\bR_\textrm{H}^{\frac{1}{2}},
\end{align}

We can rewrite \eqref{eq:Inversionmatrix} into the following form by using the block matrix inversion lemma \cite{HornMatrix1990}:
\begin{center}
\begin{math}
\bI-\bR_\textrm{H}^{\frac{1}{2}}\widetilde{\bX}^\HH\bR_1^{-1}\widetilde{\bX}\bR_\textrm{H}^{\frac{1}{2}}=(\bE_A\bC_A^{-1}\bE_A^\HH)^{-1},
\end{math}
\end{center}
where $\bE_A=[\bI_{N_{\textrm{TR}}},\textbf0_{N_{\textrm{TR}}\times N_{\textrm{RL}}}]$,
\begin{equation}
\bC_A=
\begin{bmatrix}
\bI_{N_{\textrm{TR}}}&\bR_\textrm{H}^{\frac{1}{2}}\widetilde{\bX}^\HH \\
\widetilde{\bX}\bR_\textrm{H}^{\frac{1}{2}} & \bR_1
\end{bmatrix},
\end{equation}
and $N_\textrm{TR}=N_\textrm{T}N_\textrm{R}$.%, $N_{\textrm{RL}} = L N_\textrm{R}$.

Thus, Part I of the objective function can be rewritten as:
\begin{equation}
\log\det(\bR_\textrm{H}^{\frac{1}{2}}\widetilde{\bR}_{X}\bR_\textrm{H}^{\frac{1}{2}}+\bI)=\log\det(\bE_A\bC_A^{-1}\bE_A^\HH).
\end{equation}

Noting that $\bE_A$ has full row rank, we can verify that $\log\det(\bE_A\bC_A^{-1}\bE_A^\HH)$ is convex with respect to (w.r.t.) $\bC_A$\cite{NaghshInformation2017}. In addition, since convex functions are minorized by their supporting hyperplanes \cite{BoydConvex2004}, we have
\begin{align}
\log\det(\bE_A\bC_A^{-1}\bE_A^\HH)\geq
&\log\det(\bE_A(\bC_{A,k})^{-1}\bE_A^\HH)
\nonumber\\
&+\textrm{tr}[\bT_k(\bC_A-\bC_{A,k})],
\end{align}
where $\bT_k=-\bC_{A,k}^{-1}\bE_A^\HH(\bE_A\bC_{A,k}^{-1}\bE_A^\HH)^{-1}\bE_A\bC_{A,k}^{-1}$ is the gradient of $\log\det(\bE_A\bC_A^{-1}\bE_A^\HH)$ at $\bC_{A,k}$ \cite{HjorungnesDifferentiation2007}. Then we let $\bT_k$ be partitioned as
\begin{equation}
\bT_k=
\begin{bmatrix}
\bT_{k}^{11} & \bT_{k}^{12} \\
(\bT_{k}^{12})^\HH & \bT_{k}^{22}
\end{bmatrix},
\end{equation}
where $\bT_{k}^{11}\in\mathbb{C}^{N_{\textrm{TR}}\times N_{\textrm{TR}}}$, $\bT_{k}^{12}\in\mathbb{C}^{N_{\textrm{TR}}\times N_{\textrm{RL}}}$, and $\bT_{k}^{22}\in\mathbb{C}^{N_{\textrm{RL}}\times N_{\textrm{RL}}}$. Then tr$(\bT_{k}\bC_A)$ can be written as
\begin{align*}
\textrm{tr}(\bT_k\bC_A)=
&c_{0,k}+2\Re[\textrm{tr}(\widetilde{\bX}\bR_\textrm{H}^{\frac{1}{2}}\bT_{k}^{12})]%+(\bT_{k}^{12})^\HH\bR_\textrm{H}^{\frac{1}{2}}\widetilde{\bX}^\HH]
%\nonumber\\
+\textrm{tr}(\bT_{k}^{22}\widetilde{\bX}\bR_\textrm{H}\widetilde{\bX}^\HH),
\end{align*}
where $c_{0,k}=\textrm{tr}(\bT_{k}^{11}+\bT_{k}^{22})$ is a constant not depending on $\bX$.

Therefore, a minorizer of  $\log\det(\bR_\textrm{H}^{\frac{1}{2}}\widetilde{\bR}_{X}\bR_\textrm{H}^{\frac{1}{2}}+\bI)$ is given by
\begin{equation}\label{eq:minorizerI}
  c_{1,k}+2\Re[\textrm{tr}(\widetilde{\bX}\bR_\textrm{H}^{\frac{1}{2}}\bT_{k}^{12})] + \textrm{tr}(\bT_{k}^{22}\widetilde{\bX}\bR_\textrm{H}\widetilde{\bX}^\HH),
\end{equation}
where $c_{1,k}=c_{0,k}+\log\det(\bE_A\bC_{A,k}^{-1}\bE_A^\HH)-\textrm{tr}(\bT_{k}\bC_{A,k})$.

\subsection{Minorizing Part II}
Note that
\begin{align}
\textrm{tr}(\bR_1^{-1}\widetilde{\bX}\bh_d\bh_d^\HH\widetilde{\bX}^\HH)=\textrm{tr}[(\widetilde{\bX}\bh_d)^\HH\bR_1^{-1}\widetilde{\bX}\bh_d],
\end{align}
where $\bR_1\succ\textbf0$.
According to \cite[Lemma 1]{TangConstrained2021}, $\tr(\bA^\HH \bB^{-1} \bA)$ is jointly convex w.r.t. $\bA$ and  $\bB$. Using the property of convex functions, we can obtain
\begin{align*}
\textrm{tr}[(\widetilde{\bX}\bh_d)^\HH\bR_1^{-1}\widetilde{\bX}\bh_d] %\nonumber \\
\geq& 2\textrm{Re}[\textrm{tr}((\widetilde{\bX}_{k}\bh_d)^\HH\bR_{1,k}^{-1}\widetilde{\bX}\bh_d)]\nonumber \\
&-\textrm{tr}[(\bR_{1,k}^{-1}(\widetilde{\bX}_{k}\bh_d)(\widetilde{\bX}_{k}\bh_d)^\HH\bR_{1,k}^{-1}\bR_{1}],
\end{align*}
where $\bR_{1,k} = \widetilde{\bX}_k\bR_\textrm{H}\widetilde{\bX}_k^\HH+\sigma^2\bI$.

As a result, a minorizer of Part II of the objective is given by
\begin{equation}\label{eq:minorizerII}
  c_{2,k} + \textrm{tr}(\bZ_k\widetilde{\bX}\bR_\textrm{H}\widetilde{\bX}^\HH)+2\textrm{Re}[\textrm{tr}(\widetilde{\bX}^\HH\bW_k)],
\end{equation}
where $c_{2,k} = -\textrm{tr}(\bZ_k)$, $\bZ_k=\bR_{1,k}^{-1}(\widetilde{\bX}_k\bh_d)(\widetilde{\bX}_k\bh_d)^\HH\bR_{1,k}^{-1}$, and $\bW_k=\bR_{1,k}^{-1}(\widetilde{\bX}_k\bh_d)\bh_d^\HH$.
\subsection{Minorizing Part III}
Since $\textrm{tr}(\bR_1^{-1})$ is convex w.r.t. $\bR_1$, we can obtain
\begin{align}
\textrm{tr}(\bR_1)\geq
\nonumber
&\textrm{tr}(\bR_{1,k}^{-1}) +\textrm{tr}[-\bR_{1,k}^{-2}(\widetilde{\bX}\bR_\textrm{H}\widetilde{\bX}^{\HH}-\widetilde{\bX}_k\bR_\textrm{H}\widetilde{\bX}_k^{\HH})].
\end{align}

Thus, a minorizer of $\textrm{tr}(\bR_1^{-1})$ (i.e., Part III of the objective) is given by
\begin{equation}\label{eq:minorizerIII}
  c_{3,k}-\textrm{tr}(\bR_{1,k}^{-2}\widetilde{\bX}\bR_\textrm{H}\widetilde{\bX}^\HH),
\end{equation}
where $c_{3,k}=\textrm{tr}(\bR_{1,k}^{-1})+\textrm{tr}(\bR_{1,k}^{-2}\widetilde{\bX}_k\bR_\textrm{H}\widetilde{\bX}_k^{\HH})$.
\subsection{The Minorized Problem at the $k$th Iteration}
With the results in \eqref{eq:minorizerI}, \eqref{eq:minorizerII}, and \eqref{eq:minorizerIII}, the minorized problem at the $k$th iteration can be formulated as
\begin{align} \label{eq:kthIteration}
\max_\bX \ &2\Re[\textrm{tr}(\widetilde{\bX}^\HH \bP_k)]%+(\bT_k^{12})^\HH\bR_\textrm{H}^{\frac{1}{2}}\widetilde{\bX}^\HH]
+\textrm{tr}(\bQ_k\widetilde{\bX}\bR_\textrm{H}\widetilde{\bX}^\HH)
%\nonumber \\
%&+2\textrm{Re}[\textrm{tr}(\widetilde{\bX}^\HH\bW_k]
\nonumber \\
\textrm{s.t.} &\ \textrm{tr}(\bX\bX^\HH)\leq P_t,
\end{align}
where $\bP_k = (\bT_k^{12})^\HH\bR_\textrm{H}^{\frac{1}{2}}+\bW_k$, $\bQ_k = \bT^{22}_k-\bZ_k-\bR_{1,k}^{-2}$, and we have ignored the constant terms.

By using the identities that $\textrm{tr}(\bA^\T\bB)=\textrm{vec}^\T(\bA)\textrm{vec}(\bB)$ and $\textrm{tr}(\bA\bB\bC\bD)=\textrm{vec}^\T(\bD)(\bA\otimes\bC^\T)\textrm{vec}(\bB^\T)$\cite{BernsteinTheory2009}, we can rewrite the objective of \eqref{eq:kthIteration} as
\begin{equation}
\widetilde{\bx}^\HH\widetilde{\bM}_k\widetilde{\bx}+2\textrm{Re}(\widetilde{\bx}^\HH\widetilde{\bm}_k),
\end{equation}
where $\widetilde{\bx}=\textrm{vec}(\widetilde{\bX})$,
\begin{equation}
\widetilde{\bM}_k=\bR_\textrm{H}^*\otimes \bP_k,\widetilde{\bm}_k=\textrm{vec}(\bQ_k).
\end{equation}

According to \cite{TangMinorization2018}, $\widetilde{\bx}$ is a linear function of $\bx = \vec(\bX)$, which can be written as $\widetilde{\bx} = \bB_s\bx$,
where
$\bB_s=\bE_s\otimes\bI_\textrm{L}$,
$\bE_s=[\bE_1,\bE_2,\cdots,\bE_{N_{\textrm{TR}}}]^\T$,
with $\bE_i,i=1,2,\cdots,N_{\textrm{TR}}$, denoting an $N_\textrm{T}\times N_\textrm{R}$ elementary matrix which has a unity in the $(i_r,i_c)$-{th} element and zeros in all other positions, $i_r=1+\textrm{mod}(i-1,N_\textrm{T})$, and $i_c=\lceil \frac{i}{N_\textrm{T}}\rceil$.
Then we can reformulate the optimization problem in \eqref{eq:kthIteration} as
\begin{equation} \label{eq:Optquadratic}
\max_\bx \bx^\HH\bM_k\bx+2\textrm{Re}(\bx^\HH\bm_k),\textrm{s.t.}\  \bx^\HH\bx\leq P_t,
\end{equation}
where $\bM_k=\bB_s^\HH\widetilde{\bM}_k\bB_s$, $\bm_k=\bB_s^\HH\widetilde{\bm}_k$, and we have used the fact that $\textrm{tr}(\bX\bX^\HH)=\bx^\HH\bx$.%, and we have

It can be proved that \eqref{eq:Optquadratic} is a hidden convex problem \cite{BenTalHidden1996} and can be solved via the Lagrange multipliers method. Specifically, the associated Lagrangian is given by
\begin{equation} \label{eq:Lagrangian}
L(\bx,\nu)=\bx^\HH\bM_k\bx+2\textrm{Re}(\bx^\HH\bm_k)+\nu(\bx^\HH\bx-P_t),
\end{equation}
where $\nu$ is the Lagrange multiplier associated with the constrained set.
The maximizer can be obtained by differentiating \eqref{eq:Lagrangian} w.r.t $\bx$ and setting the differentiation to zero:
\begin{equation} \label{eq:xk_1}
\bx_{k+1}=-(\bM_k+\nu_{k+1}\bI_{N_\textrm{RL}})^{-1}\bm_k,
\end{equation}
where $\nu_{k+1}$ is the solution to the following equation:
\begin{equation}
(\bm_k)^\HH(\bM_k+\nu\bI_{N_\textrm{RL}})^{-2}\bm_k=P_t.
\end{equation}

%We summarize the proposed algorithm in Algorithm 1, where $D_k$ denotes the relative entropy associated with $\bx_k$, and $\epsilon>0$ is a predefined small value.
%
%\begin{center}
%\begin{tabular}{l}
%\hline
%\textbf{Algorithm 1:} Robust waveform design for
%MIMO radar  \\detection \\
%\hline
%\textbf{Input:} $\bh_d$, $\bR_\textrm{H}$ and $P_t$.\\
%\textbf{Initialize:} $k=0,\bx^{(0)},\epsilon$.\\
%\textbf{Output:} $\bx^{\star}$\\
%\textbf{do}\\
%\ 1: Compute $\widetilde{\bM}_k$;\\%=\bR_\textrm{H}^*\otimes[\bT_k^{22}-\bZ_k-(\widetilde{\bX}_k\bR_\textrm{H}\widetilde{\bX}_k^{\HH}+\bI)^{-2}]$;\\
%\ 2: Compute $\widetilde{\bm}_k$;\\%=\textrm{vec}((\bT^{12}_k)^\HH\bR_\textrm{H}^{\frac{1}{2}}+\bW_k)$;\\
%\ 3: $\bM_k=\bB_s^\HH\widetilde{\bM}_k\bB_s$;\\
%\ 4: $\bm_k=\bB_s^\HH\widetilde{\bm}_k$;\\
%\ 5: Update $\bx_{k+1}$ by \eqref{eq:xk_1}.\\
%\ 6: $k=k+1$.\\
%\textbf{until} $|D_k-D_{k-1}|/D_k<\epsilon$.\\
%\hline
%\end{tabular}
%\end{center}

%Since MM method guarantees convergence, we can find that the relative entropy will converge to a certain value.

\section{Numerical Examples}
In this section, we provide numerical examples to verify the performance of the proposed algorithm. We consider a colocated MIMO radar with $N_\textrm{T}=6$ transmitters and $N_\textrm{R}=6$ receivers. The inter-element spacings of the transmit array and the receive array  are $d_\textrm{T}=2\lambda$ and $d_\textrm{R}=\lambda/2$, respectively ($\lambda$ is the wavelength). The code length is $L=20$. We assume that the nominal DOA of the target is $\theta_d=15^{\circ}$ (i.e., the prior knowledge). $\bh_d=\alpha_d\ba(\theta_d)\otimes \bb(\theta_d)$ with $\alpha_d=\sqrt{3/2}$ denoting the target amplitude. We model $\bR_\textrm{H}$ by $\bR_\textrm{H}=\sum_k\nolimits \sigma^2_r(\bb(\theta_k)\bb(\theta_k)^\HH)\otimes(\ba(\theta_k)\ba(\theta_k)^\HH)$, where $\sigma^2_r=0.05$,  and $\theta_k$ are uniformly distributed from $-60^{\circ}$ to $56^{\circ}$.
We initialize the proposed algorithm with randomly generated quasi-orthogonal waveforms. Finally, the proposed algorithm terminates if $|D_k-D_{k-1}|/D_k<\epsilon=10^{-4}$. %Finally, all the numerical analysis is performed on a standard laptop with AMD Ryzen 5 PRO 3500U and 16 GB of RAM.

%Before we design waveform, we assume the DOA of the deterministic scatter of the Swerling 0 target is $\theta_d=15^{\circ}$. Hence, the deterministic response $\bh_d=\alpha_d\ba_\textrm{R}(\theta_s)\otimes \ba_\textrm{T}(\theta_d)$ with $\alpha_d$ denoting the amplitude of the deterministic scatter. Then we let $\widetilde{\bR}_\textrm{H}$ be the covariance matrix of an AR(1) process, i.e. $\widetilde{\bR}_\textrm{H}(i,j)=\mu^{|i-j|}$ with $\mu=0.5$. Ignoring the spatial correlations between $\bh_n,n=1,2,\ldots,N_R$, we can obtain $\bR_\textrm{H}=\bI_{N_\textrm{R}}\otimes \widetilde{\bR}_\textrm{H}$. We initialize the proposed algorithm with randomly generated quasi-orthogonal waveforms.

\begin{figure}[!htbp]
\centerline{\includegraphics[width=8cm]{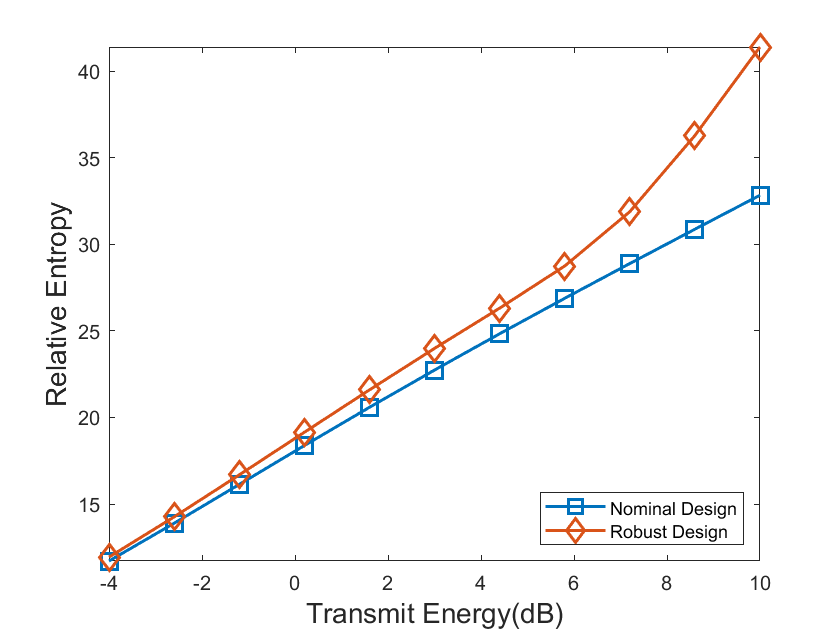}}
\caption{The relative entropy of the synthesized waveforms against transmit energy. The DOA of the target is  $\theta_{t}=25^{\circ}$.}
\label{fig1}
\end{figure}

\begin{figure}[!htbp]
\centerline{\includegraphics[width=8cm]{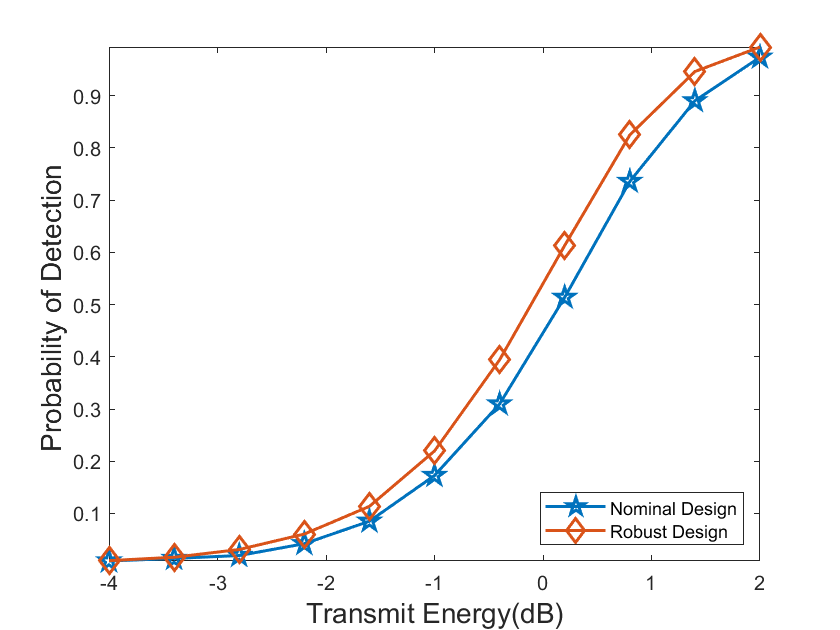}}
\caption{The probability of detection of the synthesized waveforms against transmit energy. The DOA of the target is  $\theta_{t}=25^{\circ}$.}
\label{fig2}
\end{figure}

%We made a comparison of performances between two methods. One is the proposed waveform, the other is the optimal waveform for Swerling 0 target.
Now we compare the performance  of the waveforms synthesized by the proposed algorithm with that of the waveforms synthesized by \eqref{eq:NominalSolution}.  Note that to design the waveforms by \eqref{eq:NominalSolution},  we replace $\bH$ by the nominal target response matrix $\bH_d$, and $\vec(\bH_d) = \bh_d$. Fig. \ref{fig1} shows the relative entropy of the synthesized waveforms versus different transmit energy, where the DOA of the true target is $\theta_{t}=25^{\circ}$ (i.e., a $10^\circ$ mismatch between the nominal DOA and the true DOA). We can observe that the waveforms synthesized by the proposed algorithm always have a larger relative entropy.  Fig. \ref{fig2} shows the probabilities of detection associated with Fig. \ref{fig1}, where the NP detector in \eqref{eq:NPdetector} is used to analyze the detection performance, the probability of false alarm is $P_{fa}=10^{-3}$, and $10^5$ Monte Carlo trials are conducted to obtain the threshold and the probability of detection, respectively. We can find that the results in Fig. \ref{fig2} are consistent with Fig. \ref{fig1}, showing the robustness of the proposed waveforms against the angle mismatch.

\begin{figure}[htbp]
\centerline{\includegraphics[width=8cm]{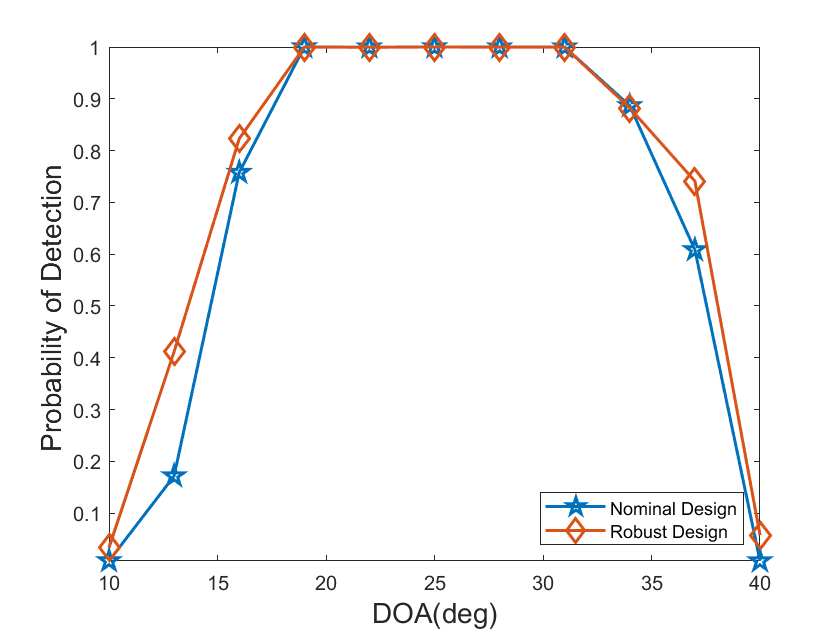}}
\caption{The probability of detection of the synthesized waveforms versus the nominal target DOA.}
\label{fig3}
\end{figure}

Fig. \ref{fig3} compares the detection probability of the proposed algorithm with that of the nominal design, where we assume that the true DOA of the target is fixed to be $25^{\circ}$, the nominal DOA of the target varies from $10^\circ$ to $40^\circ$, and the transmit energy is $P_t=1.25$. The results demonstrate that when the angle mismatch is larger than $6^\circ$, the proposed algorithm exhibits better robustness.

%In Fig. \ref{fig3}, we assume the DOA of the real target is $25^{\circ}$. Then we plot the detection performances of synthesized waveforms versus DOA of the target of prior information. We can find the proposed algorithm has higher probability of detection. Thus the robustness is shown in the result.
%Fig. \ref{fig3} shows the detection performances of synthesized waveforms versus DOA of the target, with the transmit energy $P_t=1.5$. We can find the proposed algorithm has higher probability of detection. Thus the robustness is shown in the result.

%In the examples, we assume the real DOAs of targets deviate from the prior information. In Fig 2, the DOA of the target is $\theta_{d_1}=25^{\circ}$, and In Fig 3, the DOA of the target is $\theta_{d_2}=30^{\circ}$.
%In our examples, we set the probability of false alarm $P_{fa}=10^{-3}$, then we made $10^5$ Monte Carlo trials to obtain the probability of detection.

%Because the DOA of target in the example is not the same as which we used to design waveforms, the prior knowledge is mismatched. The results show that the proposed algorithm has larger relative entropy and a superior performance of probability of detection for target. Therefore, the proposed method is efficient. It has a robust performance under the circumstance of the prior knowledge mismatch.

%\begin{figure}[!htbp]
%\centerline{\includegraphics[width=0.32\textwidth]{Fig1.png}}
%\caption{The relative entropy of the synthesized waveforms against transmit energy. The DOA of the target is  $\theta_{t}=25^{\circ}$.}
%\label{fig1}
%\end{figure}

%\begin{figure}[htbp]
%  \centering
%  \includegraphics[width=8cm]{Fig1.eps}\\
%  \caption{{Example of a figure caption.}}\label{fig-1}
%\end{figure}
\section{Conclusion}
This paper considered robust waveform design for MIMO radar target detection. A probabilistic model was proposed to describe the target uncertainty, and the relative entropy between the PDF of the observations under two hypotheses was employed as the waveform design metric. To tackle the non-convex waveform design problem, an efficient optimization algorithm based on MM was developed. Numerical results show that the waveforms synthesized by the proposed algorithm are more robust to the target mismatches.

%it adopts MM technique and is based on the principle of information theory.
%We proved the effectiveness of the proposed method. It has better performance than the optimal waveform for Swerling 0 target.
%The proposed algorithm is designed for target detection without clutter and interference. So if we consider the influence of clutter and interference, the performance of the proposed algorithm will decline. Hence in the future, we will pay more attention to the study of MIMO radar target detection in clutter and interference to get closer to reality.

\section*{Acknowledgment}
This work was supported in part by the National Natural Science Foundation of China under Grant 62171450 and 61671453, the Anhui Provincial Natural Science Foundation under Grant 2108085J30, and the Young Elite Scientist Sponsorship Program of CAST under Grant 17-JCJQ-QT-041.

%\begin{thebibliography}{00}
%\bibitem{b1} G. Eason, B. Noble, and I. N. Sneddon, ``On certain integrals of Lipschitz-Hankel type involving products of Bessel functions,'' Phil. Trans. Roy. Soc. London, vol. A247, pp. 529--551, April 1955.
%\bibitem{b2} J. Clerk Maxwell, A Treatise on Electricity and Magnetism, 3rd ed., vol. 2. Oxford: Clarendon, 1892, pp.68--73.
%\bibitem{b3} I. S. Jacobs and C. P. Bean, ``Fine particles, thin films and exchange anisotropy,'' in Magnetism, vol. III, G. T. Rado and H. Suhl, Eds. New York: Academic, 1963, pp. 271--350.
%\bibitem{b4} K. Elissa, ``Title of paper if known,'' unpublished.
%\bibitem{b5} R. Nicole, ``Title of paper with only first word capitalized,'' J. Name Stand. Abbrev., in press.
%\bibitem{b6} Y. Yorozu, M. Hirano, K. Oka, and Y. Tagawa, ``Electron spectroscopy studies on magneto-optical media and plastic substrate interface,'' IEEE Transl. J. Magn. Japan, vol. 2, pp. 740--741, August 1987 [Digests 9th Annual Conf. Magnetics Japan, p. 301, 1982].
%\bibitem{b7} M. Young, The Technical Writer's Handbook. Mill Valley, CA: University Science, 1989.
%\end{thebibliography}

% LiMIMObook2008,
\bibliographystyle{IEEEtran}
\bibliography{An_BIB}

\begin{thebibliography}{10}
\providecommand{\url}[1]{#1}
\csname url@samestyle\endcsname
\providecommand{\newblock}{\relax}
\providecommand{\bibinfo}[2]{#2}
\providecommand{\BIBentrySTDinterwordspacing}{\spaceskip=0pt\relax}
\providecommand{\BIBentryALTinterwordstretchfactor}{4}
\providecommand{\BIBentryALTinterwordspacing}{\spaceskip=\fontdimen2\font plus
\BIBentryALTinterwordstretchfactor\fontdimen3\font minus
  \fontdimen4\font\relax}
\providecommand{\BIBforeignlanguage}[2]{{%
\expandafter\ifx\csname l@#1\endcsname\relax
\typeout{** WARNING: IEEEtran.bst: No hyphenation pattern has been}%
\typeout{** loaded for the language `#1'. Using the pattern for}%
\typeout{** the default language instead.}%
\else
\language=\csname l@#1\endcsname
\fi
#2}}
\providecommand{\BIBdecl}{\relax}
\BIBdecl

\bibitem{LiMIMObook2008}
J.~Li and P.~Stoica, \emph{{MIMO} Radar Signal Processing}.\hskip 1em plus
  0.5em minus 0.4em\relax Hoboken, NJ, USA: Wiley, 2008.

\bibitem{HaimovichSeparated2008}
A.~M. Haimovich, R.~S. Blum, and L.~J. Cimini, ``{MIMO} radar with widely
  separated antennas,'' \emph{IEEE Signal Processing Magazine}, vol.~25, no.~1,
  pp. 116--129, 2008.

\bibitem{LiColocated2007}
J.~Li and P.~Stoica, ``{MIMO} radar with colocated antennas,'' \emph{IEEE
  Signal Processing Magazine}, vol.~24, no.~5, pp. 106--114, 2007.

\bibitem{HeCorrelations2009}
H.~He, P.~Stoica, and J.~Li, ``Designing unimodular sequence sets with good
  correlations-including an application to {MIMO} radar,'' \emph{IEEE
  Transactions on Signal Processing}, vol.~57, no.~11, pp. 4391--4405, 2009.

\bibitem{HeWaveformBook2012}
H.~He, J.~Li, and P.~Stoica, \emph{Waveform Design for Active Sensing Systems:
  A Computational Approach}.\hskip 1em plus 0.5em minus 0.4em\relax Cambridge,
  U.K.: Cambridge Univ. Press, 2012.

\bibitem{ChenExtended2009}
C.~Chen and P.~P. Vaidyanathan, ``{MIMO} radar waveform optimization with prior
  information of the extended target and clutter,'' \emph{IEEE Transactions on
  Signal Processing}, vol.~57, no.~9, pp. 3533--3544, 2009.

\bibitem{TangJoint2016}
B.~Tang and J.~Tang, ``Joint design of transmit waveforms and receive filters
  for {MIMO} radar space-time adaptive processing,'' \emph{IEEE Transactions on
  Signal Processing}, vol.~64, no.~18, pp. 4707--4722, 2016.

\bibitem{Cui2017SpaceTime}
G.~Cui, X.~Yu, V.~Carotenuto, and L.~Kong, ``Space-time transmit code and
  receive filter design for colocated {MIMO} radar,'' \emph{IEEE Transactions
  on Signal Processing}, vol.~65, no.~5, pp. 1116--1129, 2017.

\bibitem{Tang2020Polyphase}
B.~Tang, J.~Tuck, and P.~Stoica, ``Polyphase waveform design for {MIMO} radar
  space time adaptive processing,'' \emph{IEEE Transactions on Signal
  Processing}, vol.~68, pp. 2143--2154, 2020.

\bibitem{MaioPrinciples2007}
A.~D. Maio and M.~Lops, ``Design principles of {MIMO} radar detectors,''
  \emph{IEEE Transactions on Aerospace and Electronic Systems}, vol.~43, no.~3,
  pp. 886--898, 2007.

\bibitem{Tang2010IT}
B.~Tang, J.~Tang, and Y.~Peng, ``{MIMO} radar waveform design in colored noise
  based on information theory,'' \emph{IEEE Transactions on Signal Processing},
  vol.~58, no.~9, pp. 4684--4697, 2010.

\bibitem{Tang2019Spectrally}
B.~Tang and J.~Li, ``Spectrally constrained {MIMO} radar waveform design based
  on mutual information,'' \emph{IEEE Transactions on Signal Processing},
  vol.~67, no.~3, pp. 821--834, 2019.

\bibitem{TangEntropy2015}
B.~Tang, M.~M. Naghsh, and J.~Tang, ``Relative entropy-based waveform design
  for {MIMO} radar detection in the presence of clutter and interference,''
  \emph{IEEE Transactions on Signal Processing}, vol.~63, no.~14, pp.
  3783--3796, 2015.

\bibitem{Tang2021RangeSpread}
B.~Tang and P.~Stoica, ``Information-theoretic waveform design for {MIMO} radar
  detection in range-spread clutter,'' \emph{Signal Processing}, vol. 182, p.
  107961, 2021.

\bibitem{Tang2016Rician}
B.~Tang, J.~Tang, and Y.~Zhang, ``Design of multiple-input-multiple-output
  radar waveforms for {Rician} target detection,'' \emph{IET Radar, Sonar \&
  Navigation}, vol.~10, no.~9, pp. 1583--1593, 2016.

\bibitem{TangMinorization2018}
B.~Tang, Y.~Zhang, and J.~Tang, ``An efficient minorization maximization
  approach for {MIMO} radar waveform optimization via relative entropy,''
  \emph{IEEE Transactions on Signal Processing}, vol.~66, no.~2, pp. 400--411,
  2018.

\bibitem{Kaystatistical1998}
S.~M. Kay, \emph{Fundamentals of statistical signal processing, vol. ii:
  detection theory}, Prentice-Hall, Upper Saddle River, Newer Jersey, 1998.

\bibitem{HornMatrix1990}
R.~A. Horn and C.~R. Jonson, \emph{Matrix Analysis}.\hskip 1em plus 0.5em minus
  0.4em\relax Cambridge, U.K: Cambridge Univ. Press, 1990.

\bibitem{NaghshInformation2017}
M.~M. Naghsh, M.~Modarres-Hashemi, M.~A. Kerahroodi, and E.~H.~M. Alian, ``An
  information theoretic approach to robust constrained code design for {MIMO}
  radars,'' \emph{IEEE Transactions on Signal Processing}, vol.~65, no.~14, pp.
  3647--3661, 2017.

\bibitem{BoydConvex2004}
S.~Boyd and L.~Vandenberghe, \emph{Convex Optimization}.\hskip 1em plus 0.5em
  minus 0.4em\relax Cambridge, U.K: Combridge Univ. Press, 2004.

\bibitem{HjorungnesDifferentiation2007}
A.~Hjorungnes and D.~Gesbert, ``Complex-valued matrix differentiation:
  Techniques and key results,'' \emph{IEEE Transactions on Signal Processing},
  vol.~55, no.~6, pp. 2740--2746, 2007.

\bibitem{TangConstrained2021}
B.~Tang, J.~Liu, H.~Wang, and Y.~Hu, ``Constrained radar waveform design for
  range profiling,'' \emph{IEEE Transactions on Signal Processing}, vol.~69,
  pp. 1924--1937, 2021.

\bibitem{BernsteinTheory2009}
D.~S. Bernstein, \emph{Matrix Mathematics: Theory, Facts, and Formulas}.\hskip
  1em plus 0.5em minus 0.4em\relax Princeton, NJ, USA: Princeton Univ. Press,
  2009.

\bibitem{BenTalHidden1996}
A.~Ben-Tal and M.~Teboulle, ``Hidden convexity in some nonconvex quadratically
  constrained quadratic programming,'' \emph{Math. Program.}, vol.~72, no.~1,
  pp. 52--63, 1996.

\end{thebibliography}
\end{document}